\documentclass[12pt]{article}
\input epsf.sty
\def\be{\begin{equation}} \def\ee{\end{equation}}
\def\bea{\begin{eqnarray}} \def\eea{\end{eqnarray}} \def\ba{\begin{array}}
\def\ea{\end{array}} \def\ben{\begin{enumerate}} \def\een{\end{enumerate}}
  
\def\lll{\label}
\newcommand{\eqn}[1]{(\ref{#1})}

\newcommand{\plb}[3]{Phys. Lett. {\bf B#1} ({#2}) {#3}}
\newcommand{\prd}[3]{Phys. Rev. {\bf D#1} ({#2}) {#3}}
\newcommand{\hepth}[1]{{\tt hep-th/{#1}}}

\def\ov{\over}
\def\br{\nonumber\\}

\begin{document}
{}~
\hfill\vbox{\hbox{hep-th/0505012}\hbox{SINP-05/08}}\break

\vskip .5cm
\centerline{\Large \bf
More on Tachyon Cosmology in De Sitter Gravity}

\vskip .5cm

\vspace*{.5cm}

\centerline{\large \sc Harvendra Singh}

\vspace*{.5cm}

\centerline{\large \it Theory Division, Saha Institute of Nuclear 
Physics}

\centerline{\large \it  1-AF Bidhannagar,
Kolkata-700064, INDIA}

\centerline{E-mail: h.singh@saha.ac.in}

\vskip1cm
\centerline{\bf Abstract} \bigskip

We aim to study rolling tachyon cosmological solutions in 
de Sitter gravity. The solutions  are taken to be 
flat FRW type and these are not time-reversal symmetric. 
We find that cosmological constant of our universe has to be fine-tuned 
at the level of the action itself, as in KKLT string compactification. 
The rolling tachyon can give rise to 
required inflation with suitable choice of the initial conditions
which include nonvanishing Hubble constant. We also determine an upper 
bound on the volume of the compactification manifold.

\vfill \eject

\baselineskip=16.2pt

\section{Introduction}
Recently the phenomenon of tachyon condensation \cite{sen1,sen2}  
has been a subject of much attention in string theory as well as in 
cosmology \cite{sen3,linde3}. The tachyon field appears as 
an instability in non-supersymmetric 
$p$-branes dynamics as well as an instability in brane-anti-brane system 
in superstring theory. The low energy 
tachyonic field theory is governed by
the Born-Infeld type action
\cite{sen0,sen1,sen2}
\be\lll{tachyact}
-\int d^4x V(T)\sqrt{-{\rm det} (g_{\mu\nu}+2\pi\alpha '\partial_\mu 
T\partial_\nu 
T)} \ee
where tachyon $T$ appears explicitly as a world volume scalar 
field, 
$g_{\mu\nu}$ represents the pull-back of the spacetime metric and $V(T)$ 
is a positive definite tachyon potential. In the flat spacetime equation 
of motion of this action, 
for purely time-dependent tachyon, assume following unique form:
\be\lll{a1}
{V(T)\over\sqrt{1-2\pi\alpha '\dot T^2}}=\rho.
\ee
It means that the tachyon field behaves like a fluid of 
constant positive energy 
density, $\rho$, and of negative pressure, $P=-V\sqrt{1-2\pi\alpha '\dot T^2}$. During 
the time evolution tachyon field rolls down the potential and reaches its 
true vacuum value $T=\infty$ where $V(\infty)=0$, while keeping $\rho$ 
constant. When this tachyon system is 
coupled to background supergravity 
this result will, however, get significantly modified and new type of 
solutions can be  obtained. In recent time   
the applications of open string tachyon in inflationary 
cosmological models have been 
quite encouraging  
\cite{gibbon,faityt,cgjs,linde,sami,peet,roy,odintsov, 
kklt2,garousi,sen3,linde3,linde2}.
\footnote{
Also a wide class of useful references on this topic can be found in 
\cite{sen3,linde3,paddy}.}

In this paper we plan to study a very specific model in which tachyon 
field is coupled to 
de Sitter gravity. It will be assumed that compactification of string 
theory to four-dimensional de Sitter spacetime of KKLT type \cite{kklt} 
can be achieved in which all 
the moduli including the volume modulus are fixed. Our aim is  
to study spatially 
flat, homogeneous cosmological solutions in this model. The solutions we
discuss are completely {\it non-singular} and could describe inflationary 
situations just like after the big-bang and also at late time may 
possibly describe  
the universe as of today. However, it will require fine-tuning of the 
cosmological constant at the supergravity level itself.   
The paper is organised as follows. In section-2 and section-3 we work out 
the salient 
features of simple inflationary models which are obtained by  
coupling bulk supergravity to the tachyon action. In 
section-4, we  
consider the case where gravity with bulk cosmological constant is coupled 
to the tachyon action. We  restrict only to small 
cosmological 
constant so that its effects on tachyon evolution are minimal. We 
 estimate the  number of e-folds during 
inflation, and discuss the naturalness problem in the context of smallness 
of the observed cosmological constant. We also obtain an upper 
bound on the size of compactification manifold. Section-5 contains the 
numerical analysis where we can explicitly see the behaviour of various 
physical quantities. We summarise the main results in 
section-6.

\section{ De Sitter gravity with tachyon field}
 
We follow the model considered by Sen \cite{sen3}  based on tachyon
field theory coupled to background gravity and bulk cosmological constant,
$\Lambda$. We 
assume that all moduli are frozen possibly as in KKLT scenario
\cite{kklt}.  The
four-dimensional effective action is
taken as
\be\lll{1a}
S=\int d^4x \bigg[ {1\over 16\pi G}\sqrt{-g}( R 
-2\Lambda)-V(T)\sqrt{-{\rm det} (g_{\mu\nu}+\partial_\mu T\partial_\nu T)}
\bigg]
\ee
where Newton's constant $G$ is related to   
four-dimensional Planck mass $M_p$ as $8\pi G= M_p^{-2}$. We 
have also 
taken $2\pi\alpha '=1$ which sets string tension to unity. In this section
our analysis
will not explicitly depend upon the actual form of $V(T)$. But, we
shall require that at the top of the potential, $V(T_0)=V_0$ is 
large but finite, while $V(\infty)=0$ in the vacuum. These consist two 
well known properties of the tachyon potential, e.g.
$V(T)=V_0/{\rm cosh}(T/\sqrt{2})$ in superstring theory \cite{sen2}.
Later on we
shall take $\Lambda\ll G V_0 $, because we would like to have  solutions 
in our 
model to emerge with the properties of the universe as observed today, and 
also with this condition effect of $\Lambda$ on tachyon 
rolling will be minimal. 
Motivated by experimental inputs,
that universe appears {\it spatially flat} and homogeneous at large 
scales, 
we shall look for time-dependent but not necessarily 
time-reversal symmetric solutions of
equations of motion of the
 action.\footnote{Time reversal symmetric solutions have been discussed by 
Sen \cite{sen3}. The recent cosmological measurements (WMAP) could 
be found in \cite{wmap}.} 
Hence we take the metric ansatz to be
 flat Friedman-Robertson-Walker type
\be\label{1a1}
ds^2=-dt^2 +a(t)^2\left(dr^2+r^2(d\theta^2+sin^2\theta d\phi^2)\right)
\ee
and the tachyon to be purely time-dependent $T=T(t)$.

With the above ansatze  
the field equations derived from the action 
can be written as, the Tachyon equation:
\be\lll{teqn}
\ddot T=-(1-\dot T^2)\left({V'\over V}+3\dot T {\dot a\over a}\right) 
\ee
the Friedman equation:
\be\lll{feqn}
{\ddot a \over a}= {\Lambda\ov 3}+{8\pi G\ov
3}\bigg[{V(T)\over\sqrt{1-\dot T^2}}(1 -{3\ov2} \dot T^2)\bigg] 
\ee
and the Raychaudhuri equation:
\be\lll{reqn}
\left({\dot a\over a}\right)^2={\Lambda\ov 3}+{8\pi G\ov
3}{V(T)\over\sqrt{1-\dot T^2}}.
\ee
It can be seen that the equation \eqn{teqn} 
follows simply by taking the time derivative of the eq. \eqn{reqn}. 
Therefore equations \eqn{feqn} and \eqn{reqn} contain all information 
about tachyon rolling in this model, making \eqn{teqn} redundant. 
Let us introduce the quantity $H(t)\equiv\dot a(t)/a(t)$  known as Hubble
constant at any given time $t$. In this simple model equations
\eqn{feqn} and \eqn{reqn}
immediately give us 
\be\lll{keqn}
\dot H=-{3\over 2} {8\pi G\ov
3}{V(T)\over\sqrt{1-\dot T^2}} \dot T^2
\ee
This is the key equations for the evolution of the Hubble constant.
Since the quantity ${V(T)\over\sqrt{1-\dot T^2}}$ on the right hand side 
of the 
equation \eqn{keqn} is always positive definite, 
while $\dot T^2$
varying between 0 and 1 monotonically, this equation 
implies, 
$${dH\over dt}\le0.$$ 
That is, {\it no matter where we are in
tachyon evolution, 
Hubble constant always decreases with time  or at most becomes
constant, but it cannot increase in time.} 
We shall see explicitly in the following that $H(\infty)$ is either 
vanishing 
or assumes a constant value at late times.

The equations \eqn{reqn} and the \eqn{keqn} can further be assembled in
the following useful form
\be\lll{ieqn}
{\dot H\over H^2-\lambda^2}=-{3\over 2}\dot T^2
\ee
where $\lambda^2\equiv\Lambda/3$. Note that from \eqn{reqn} only those
situations are allowed for which $H(t)\ge\lambda$. This equation can be
immediately
integrated to know the actual time dependence of  $H$ provided
we know how $\dot T$ changes with time, for that one has to solve tachyon
equation \eqn{teqn} exactly for the given potential $V$. Nevertheless 
many useful conclusions can still be drawn. We will consider
two special
cases where  $\Lambda=0$ and  $\Lambda> 0$ in what follows.

\section{Cosmology with $\Lambda=0$}

This case has rather been studied  in quite 
detail in the literature \cite{gibbon,faityt,cgjs,sami}.\footnote{ 
To  collect all important  references on this, however, one 
can see recent 
papers \cite{sen3,linde3} which include most of them.} We study this case 
here again to gain more insight. In fact
in this particular case \eqn{ieqn} reduces to a very simple form 
\be\lll{ieqn1}
{\dot H\over H^2}=-{3\over 2}\dot T^2~~~~{\rm Or}~~~~
{d\over dt}({ H}^{-1})={3\over 2}\dot T^2
\ee
which can be integrated immediately to give {\it Hubble time} (the inverse 
of Hubble 
constant)
\be\lll{ieqn1a}
H^{-1}(t)={3\over2}\int_0^t\dot T^2(t') dt' + {1\over H_0}\equiv F(t)
\ee
where $H_0$ is an integration constant which is positive definite and has 
to be fixed from 
initial
conditions at $t=0$ . The usefulness of the above equation is
evident
in that it is an integration of the function  $(\dot T)^2$ over the time
interval $t$ elapsed after $t=0$. Since $\dot T^2$ is a well behaved
quantity and can only vary between 0 and 1, 
without knowing its exact form we can
draw following conclusions;

(1) As $t\to\infty$, we will have $F(t)\to \infty$ and so
$H(\infty)=
0$. Thus Hubble constant vanishes in future and no-matter-what universe
becomes flat.

(2) Consider two different times $t_1,~ t_2$ ($t_1<t_2$), from 
\eqn{ieqn1a} the change 
in Hubble time 
in the interval $\triangle t=t_2-t_1$ is given by
\be\lll{ieqn1b}
\triangle 
H^{-1}=H^{-1}(t_2)-H^{-1}(t_1)={3\over2}\int_{t_1}^{t_2}\dot 
T^2(t) dt
\ee
which is independent of the initial value $H_0$. If $\dot 
T^2 $ varies slowly in the short interval $\triangle t$ then
\be\lll{ieqn1c}
{\triangle 
H^{-1}\over\triangle 
t}\simeq{3\over2}\dot T^2(t_1)
\ee
Thus by measuring the Hubble constant at two different instances will give
us an idea of rate of change of tachyon field and vice-versa. Also since 
$\dot T^2\le 1$ this provides us with a bound, that for a short interval 
$\triangle 
t$
\be\lll{ieqn1cc}
{\triangle 
H^{-1}\over\triangle 
t}\le{3\over2}.
\ee
which can be easily tested.   
Now coming back, the Raychaudhuri equation \eqn{reqn}  becomes
\be\lll{ieqn1aa}
{8\pi G\ov
3}{V(T)\over\sqrt{1-\dot T^2}}=H^2=\left({1\ov F(t)}\right)^2
\ee
We see that tachyon equation \eqn{teqn} is immediately satisfied by 
\eqn{ieqn1aa}. Thus we
have been able to reduce the problem to the first derivatives of the
quantities involved. Now we can
fix other things from here. Since
$F(\infty)=\infty$ we learn from \eqn{ieqn1aa} 
${V(\infty)\over\sqrt{1-\dot T^2}}= 0$ at $t=\infty$. It means
as time passes, $V\to0$ faster than $\dot T\to1$ so that 
${V(T)\over\sqrt{1-\dot T^2}}\to 0$. More precisely, it is evident from
\eqn{ieqn1a} that at late stage in the evolution $H(t)\propto {1\over t}$ 
that implies 
$${\rm Lim}_{t\to\infty}{V(T)\over\sqrt{1-\dot T^2}}\propto {1\over 
t^2}.$$ 
This is 
quite unlike in pure tachyon condensation where 
$V\to0$ with  
$\dot T\to1$, keeping ${V(T)\over\sqrt{1-\dot T^2}}$ fixed, see eq. 
\eqn{a1}. 

Now at $t=0$, $T(0)=T_0=0$ 
and $V=V_0$,
but what about the initial value of $\dot T$ at $t=0$? We find that $\dot 
T(0)$ can have  any
generic value between 0 and 1, but certainly it must be less than 1. It is 
plausible to take $\dot T(0)=0$, with that
\be
 H(0)=H_0=\sqrt{{8\pi G\over 3}V_0},
\ee
which makes initial value of the Hubble constant  tuned to the height 
of tachyon potential. Also
this initial value has other important implications for inflation in our 
model. 
With the choice that $\dot T(0)=0$, we immediately learn from eq.
\eqn{ieqn1} that near the top of the potential 
$$|\dot H|\ll H^2$$
automatically, which is the key requirement for inflation in the 
cosmological models.\footnote{ To emphasize this inflationary 
condition directly follows when $\dot T \sim 0$ and has no bearing on the 
value of $H$. In 
fact inflation 
lasts so long as $\dot T^2\ll {2/3}$.}   
Also if we want the initial value
 $H_0$ to be large, only way we can do it is by increasing the height
$V_0$ of the tachyon potential. Taking $H_0$ very large but finite,
an initial situation we may describe as just after 
{\it big-bang}, where universe starts with an explosion.~\footnote{ 
However there 
are problems in taking $V_0$ large in string tachyon models, see 
\cite{linde}. 
To estimate,  $GV_0\sim {N g_sM_s^2\over 
v_0}$, where  $g_s\ll1$ is string coupling, 
$v_0\equiv(R/l_s)^6$ is the dimensionless compactification volume 
parameter and  $N$ is the number of coincident D3-branes. 
 When converted to four-dimensional Planck mass units, using the relation
$M_s=\sqrt{g_s^2\over v_0} M_p$, we find
 $GV_0\sim{ N g_s^3\over 
v_0^2}M_p^2$ is, in general, astronomically  much larger than currently 
observed value of cosmological constant,
$\Lambda_{observed}\approx 10^{-122} M_p^2$. 
}. 
It should  however be emphasized that our solutions are completely {\it 
non singular}. Note that 
we are dealing with an effective action and so initial big-bang 
singularity, if any, cannot be addressed
in this model. For that one will have to work in full string theory. 

In summary, we find from eqn \eqn{ieqn1a} that in this $\Lambda=0$ model 
not 
only $\dot
H$ vanishes in
future but also $H$. That means effective cosmological constant
vanishes at late time and universe becomes flat. Before closing this 
section let us also find out how does the
tachyon evolve in its vacuum as $t\to\infty$.
In the tachyon vacuum, $ T\sim \infty, H\sim 0$ while  $\dot T$ 
is still less than 1. 
The tachyon equation effectively assumes the
following form:
\be\lll{teqn1}
\ddot T\simeq(1-\dot T^2)\left({1\over \sqrt{2}}\right) 
\ee
where we have specifically taken the potential to be 
$V=V_0/{\rm cosh}(T/\sqrt{2})$. This
equation suggests that
late time evolution of tachyon is decoupled from gravity. The late 
time equation
\eqn{teqn1} has an exact solution $T\sim\sqrt{2} {\rm ln~
cosh}({t\ov\sqrt{2}})$. Thus ultimately $T\sim t$ as $t\to \infty$, i.e. 
tachyon field becomes
time or in other words $\dot T\to 1$. This has been the most common
feature in tachyon
condensation \cite{sen1,sen2}. 

\subsection{Matter dominated phase}
We can also see universal behavior as $t\to\infty$. 
In the neighbourhood of $\dot T \approx 1$,
from \eqn{ieqn1} 
$${d\over dt} H^{-1}\approx 3/2 .$$ 
This tells us that in the far future
the scale factor behaves as $a(t)\sim t^{2/3}$ and $H\sim 
{2\over3}{1\over t}$, 
which is characterised as the {\it  matter dominated
phase of universe} and is the decelerating one ($\ddot a<0$) 
\cite{hartle}. 
 It means that the expanding universe enters into a decelerating 
matter dominated phase towards the end of evolution. 
The deceleration is maximum as we reach closer to 
the tachyon vacuum. Here it is useful to mention that current phase of 
our universe is a mix of matter and vacuum dominated phases.  
On that basis the models with $\Lambda=0$  perhaps can only describe the 
inflationary situations (just after big-bang) and may not 
physically describe the universe  today. 
In the next section we will see that by allowing non-zero $\Lambda$ in 
the action can change the situation.

\section{Cosmology with $\Lambda\ne 0$}

This is an important case to study and we shall assume a 
non-vanishing 
positive cosmological 
constant in the effective action \eqn{1a}, which can arise in string  
compactifications to four dimensions in very
special situations, as in KKLT compactification \cite{kklt}. Recently 
Sen has studied this type of tachyon cosmologies in \cite{sen3}. The 
difference here is, since we will consider 
only  flat FRW cosmologies, we cannot start with the  
time-reversal symmetric initial 
conditions as in \cite{sen3}. The crucial difference is the Hubble 
constant will never be zero in our model.    
We repeat the calculations of the last section but now including
the finite cosmological constant $\Lambda$. Consider the
equation \eqn{ieqn}
\be\lll{ieqn2}
{\dot H\over H^2-\lambda^2}=-{3\over 2}\dot T^2
\ee
which after integration gives
\bea\lll{ieqn2a}
H(t)=\lambda~ {\rm coth} [\lambda F(t)]
\eea
where $$ F(t)={3\over2}\int_0^t\dot T^2(t') dt' + H_0^{-1}
$$ with $H_0$ being an integration constant which has to be fixed from 
initial
conditions at $t=0$. We shall consider those situations only in which 
$H(0)\gg\lambda$. Again the usefulness of 
the above equation is evident
in the simplicity of the right hand side expression which involves area 
under the curve
$\dot T^2(t)$ over the time
interval $t$ elapsed after $t=0$. We 
 draw the following conclusion.

{\it Since for $t\to\infty$, $F(t)\to \infty$, hence
$H(\infty)=
\lambda$. Thus in this de-Sitter type model, Hubble constant becomes 
constant 
in future, no-matter
how large initial value of $H$ the universe starts with.}

The equation \eqn{reqn} becomes
\be\lll{ieqn2aa}
{8\pi G\ov
3}{V(T)\over\sqrt{1-\dot T^2}}=
\lambda^2~ {\rm cosec h}^2 [\lambda F(t)]
\ee
The tachyon equation \eqn{teqn} is immediately satisfied
by this equation.
Let us fix rest of the things from here. Since
$F(\infty)=\infty$ we learn from \eqn{ieqn2aa} that $V$ must be in the
vacuum $V(\infty)=0$ at $T=\infty$, but still $\dot T<1$. More precisely, 
we can determine from the above that, as $t\to\infty$, 
$${V(T)\over\sqrt{1-\dot T^2}}\propto 
e^{-3\lambda t}.$$

At the  
time $t=0$, $V=V_0$,  
$T(0)=T_0=0$, and with the condition
 $\dot T(0)=0$, we find that
 \be
H(0)=\lambda{\rm coth}({\lambda\over H_0} )=\sqrt{{8\pi G\over 
3}V_0+\lambda^2}.
\ee 
In this way all integration constants are fixed and initial value of
Hubble constant is fine tuned to $V_0$. Since we are taking
$\Lambda\ll 8\pi G V_0$, again only by choosing
large $V_0$
we can start with high value of Hubble constant.  

\subsection{Vacuum dominated phase}

In the tachyon vacuum, where $ T\sim \infty$, 
and $H\sim\lambda$, but still $\dot T\ne1$, the  equations of motion will 
assume following simple form:
\bea\lll{teqn2}
&&\ddot T=(1-\dot T^2)\left({1\over \sqrt{2}}-3\lambda\dot T\right) 
\br && {\dot a\over a}=\lambda
\eea
where  the potential has been $V=V_0/{\rm cosh}(T/\sqrt{2})$. This
equation suggests
 that
late time evolution of tachyon is not completely decoupled from gravity.
The tachyon equation \eqn{teqn2} can be exactly integrated to give 
\cite{ryzhik}
\be\lll{teqn3}
\left({(1-\beta \dot T)^{2}\over (1-\dot T^2)}\right)^\beta
{1+\dot T\over 1-\dot T}=e^{\sqrt{2}(1-\beta^2)(t+t_0)}
\ee
where $\beta=3\sqrt{2}\lambda$ and $t_0$ is integration constant. Thus, if 
$\beta<1$, the $t\to\infty$ limit 
of the \eqn{teqn3} suggests that $(1-\dot T)\to0$, that is $T\sim t$ 
and tachyon ultimately becomes the time. While if $\beta>1$, the 
$t\to\infty$ limit 
of the \eqn{teqn3} suggests that $(1-\beta\dot T)\to0$, that is $T\sim 
{t\over\beta}$ and tachyon becomes time only upto a factor. In other words 
$\dot T$ 
never reaches unity. Specially for $\beta=1$, i.e. $\lambda={1\over 
3\sqrt{2}}$, the integration of the \eqn{teqn2} gives
\be
{1\over 1-\dot T} + ln \sqrt{1+\dot T\over 1-\dot T} =\sqrt{2}(t +t_0).
\ee
This equation also suggests that $\dot T \sim 1$ at late time. Out of 
these possibilities, $\beta<1$ case is more favored. In a rather shorter 
analysis, 
for $\lambda\ll{1\over 3\sqrt{2}}$ the
tachyon equation effectively
\eqn{teqn2} becomes  
$$\ddot T\simeq(1-\dot T^2)\left({1\over \sqrt{2}}\right) $$ and it
has the solution $T\sim\sqrt{2} {\rm ln~
cosh}({t\ov\sqrt{2}})$. Thus ultimately $T\sim t$, i.e. tachyon becomes
time and $\dot T\to 1$. 

The scale factor corresponding to \eqn{teqn2} behaves as 
$$a(t)\sim e^{\lambda t} $$
which is like  in {\it vacuum dominated
inflationary scenario} where the universe accelerates, see \cite{hartle}. 
From \eqn{ieqn2}, during the 
evolution where $H(t)\gg\lambda$, the 
equation reduces to that of $\Lambda=0$ case, which has a matter dominated 
decelerating phase characterised by scale factor $a(t)\sim t^{2/3}$. Once 
$H\sim\lambda$ the vacuum dominated phase takes over where the scale 
factor 
behaves as $a(t)\sim e^{\lambda t} $. It means in this model the matter 
dominated phase will precede
the vacuum dominated phase during evolution. 

In summary, the universe at 
$t=0$ starts with accelerating expansion then enters into decelerating 
matter dominated intermediate phase and finally exiting into vacuum 
dominated 
accelerating phase.   
In fact from eqn. \eqn{ieqn2a} 
 $H$ reaches the terminal value $\lambda=\sqrt{\Lambda/3}$. 
It is therefore clear
that if we want $\Lambda$ to be the 
cosmological constant as observed today, 
then $\Lambda$  must
 be fine-tuned in the effective action to the observed value 
possibly through some 
stringy mechanism, something like we encounter in KKLT scenario 
\cite{kklt}.

\subsection{Number of e-folds} 
The number of e-foldings through which universe expands during the 
inflation is defined as
\be
N_e={\rm ln}{a(t_2)\over a(t_1)}\equiv\int_{t_1}^{t_2} H(t)dt.
\ee
This is an important parameter in cosmology and the estimate is that 
our universe has gone through close to $60$ e-foldings during inflation.
We would like to estimate the value of this number in tachyon cosmology. 
Consider we are at the top of the potential, in this neighborhood the 
typical value of $\dot T^2$  is infinitesimally small. Then 
from eq. \eqn{ieqn2a},
during the inflation  
Hubble constant behaves as (since ${\lambda\over H_0} \ll 1$)
$$H=\lambda {\rm coth}[\lambda({3\over2}\int_0^{t}\dot T^2 
dt'+H_0^{-1})]\simeq 
{1\over 
{3\over2}\int_0^{t}\dot T^2 dt'+H_0^{-1}}.$$
Hence the inflation lasts so long as ${3\over2}\int_0^t\dot T^2 dt' 
\ll {1\over H_0}$.    
The estimate of the number of e-folds during the inflationary interval, 
$\tau$, can be obtained also by using equation \eqn{ieqn1aa} which tells 
us
that ${8\pi GV\over3}\le H^2$ always. 
 We have an important bound
 \bea\lll{inte1}
N_e=\int_0^\tau H dt \ge 
\sqrt{{8\pi G \over3}}\int_0^{\tau} V^{1\over2} dt 
\eea
In order that we get some 
estimate of the value of $N_e$ we need to further evaluate the end 
integral in the inequality \eqn{inte1}. But it is difficult until we have 
exact 
solutions. 
Note that during inflation both $V$ 
and $H$ change in time, but from eq.\eqn{ieqn1aa} it can be 
determined that $H$ varies less slowly than $V$. So it 
can be
safely assumed that $H$ to be constant at $H\sim H_0$ though out 
inflation. 
We set $H=H_0$ 
in the integral \eqn{inte1} and assuming that inflation ends at 
$T(\tau)\simeq 1$ 
we can evaluate 
\bea
\sqrt{{8\pi G \over3}}\int_0^{\tau} V^{1\over2} dt 
&=&\int_{T(0)}^{T(\tau)}\bar V^{1\over2}{dt\over dT} dT
=\int_{T_0}^{1} \bar V^{1\over2}\left(1-{\bar 
V^2\over 
H^4}\right)^{-{1\over2}} dT \br &&\approx
\int_{T_0}^1\bar V^{1\over2} \left(1-{1\over 
{\rm cosh}^2({T\over\sqrt{2}})}\right)^{-{1\over2}} dT 
\eea
where $\bar V\equiv {8\pi G \over3} V$.
Making other crude but suitable approximations we estimate the value of 
the above integral as $\sqrt{2}H_0 {\rm ln} {1\over T_0}$, hence  
\be\lll{ne1}
N_e\ge \sqrt{2}H_0 {\rm ln} {1\over T_0}.\ee
Typically taking $H_0=1$ and $T_0=10^{-10}$ in the above, gives us 
 $N_e\ge
32.5$. The value  
$H_0\sim 1 $ would mean $H_0\sim O(1) M_p$. However, the form of 
equation \eqn{ne1} suggests that various choices of $(H_0,T_0)$
can give rise to the same value of $N_e$ and so there is flexibility in 
chosing the initial values.  In 
the next section we will also present a numerical estimates of these
quantities. 

\subsection{Naturalness and $\Lambda_{observed}$}

In this work we have been interested in taking 
$\Lambda=\Lambda_{observed}\simeq10^{-122} 
M_p^2$. However, there are a few important points left to explain here.
Let us make a theoretically more favored choice of the mass scales and 
set $M_s>M_p$. Since the two scales are related as
$M_s={g_s\over\sqrt{ 
v_0}}M_p$, one has to 
choose compactification volume smaller such that
$ {g_s^2\over v_0}>1~(g_s\ll1)$. As  
$H_0^2={8\pi GV_0\over3}\sim 
{N\over g_s}({g_s^2\over v_0})^2 M_p^2$, one finds that $H_0^2\gg 
O(1) M_p^2$, while note that $H_0\sim O(1) M_p$
corresponds to 
the Planck mass density ($\rho_{pl}\sim 10^{94} g/cm^3$) expected 
at the big bang in the inflationary models.
Thus $M_s>M_p$  makes the initial value $H_0^2$ astronomically larger 
than the 
observed value of cosmological constant in the effective action.
So apparently there is a question of naturalness  here, for example,  
in any analysis based on an effective action it 
is rather good to deal with
the range of values of a physical quantity not too far 
separated. A quantity having $10^{60}$ orders of magnitude 
difference, as is the case with Hubble constant $H(0)=H_0$ 
and $H(\infty)=\lambda$  can have problems with 
effective action \eqn{a1}. 
So there seems to be an immediate 
hierarchy (naturalness)  problem. 
On the other hand, there would have been
no problem were $H_0^2$ taken in the neighborhood of 
$\Lambda_{observed}$. This could be simply  achieved by taking large 
volume 
compactification $v_0>1$.

So let us consider the opposite situation where $M_s \ll M_p$. This  
requires large volume compactification,
that is, we must have $ {g_s^2\over v_0}\ll1~(g_s\ll1)$. 
But we need to maintain $H_0^2={8\pi GV_0\over3}\gg\Lambda$ at the same 
time 
for our 
model to work. 
This restriction translates into the bound 
$$1<{v_0\over g_s^2}\le 10^{60}.$$
So there is an upper bound on the largeness of the 
compactification volume given by
$$v_0=(R/l_s)^6\le g_s^2 10^{60}$$
beyond that our analysis breaks down. 
This implies there is a lot of freedom in choosing the size of 
compactification manifold.      
For example, if 
we consider large volume compactification saturating the bound   
${g_s^2\over {v_0}}= 10^{-60}$ with weak string coupling, 
the string scale is pushed close to  
$M_s\sim 10^{-30}M_p$.~\footnote{ Such cases 
have been discussed previously in literature  
\cite{dimarka} though entirely in different context.} That is, by taking 
the string scale $M_s$ as low 
as {\it millimeter} scale the problem of hierarchy could be immediately 
taken care of.
But this amounts to bringing down the height of tachyon 
(inflaton) potential to a 
low value.  
 So what value of $v_0$ is optimum and is of phenomenological 
 importance, will depend upon a particular model.

\section{Numerical analysis of the tachyon rolling}

We mainly aim here to reproduce the numerical results of the 
tachyon-gravity  system for $\Lambda=0$, while for $\Lambda\ne0$ the basic 
results 
will be similar except at the late time. The two key first order equations 
are
\bea\lll{aeqn1}
&&{\dot H\over H^2}=-{3\over 2}\dot T^2 \br
&&{8\pi G\ov
3}{V(T)\over\sqrt{1-\dot T^2}}=H^2
\eea
We are interested in the initial conditions where $ T(0)=0,~ H_0\ne0$
such that at the top of the potential $\dot T(0)=0$. This implies $\dot 
H(0)=0$ with $  
{8\pi G\ov
3}V(0)=H_0^2$. Classically the system with these initial conditions will 
not evolve in time, however a 
small quantum fluctuation will dislocate the configuration from the top 
and the system will start evolving. We shall take initial conditions in 
our numerical analysis in conformity with this fact where we are slightly 
away 
from the top position. The potential considered in these 
calculations is $V=V_0/Cosh({T\over\sqrt{2}})$.  
In our units $t$ is measured in $M_p^{-1}$ units and 
the Hubble constant $H$ is measured in $M_p$. It must be kept in mind that
we have already set $\alpha'=M_s^{-2}=1$ in the beginning. Now we  
also set $M_p =1$, with
 ${g_s^2\over v_0}= 1$. Note that this has been done primarily for 
simplification and it will allow us to present 
 the rolling behaviour of the equations \eqn{aeqn1}. The results 
could be exploited in a suitable phenominological setting.

\begin{figure}[!ht]
\leavevmode
\begin{center}
\epsfysize=5cm
\epsfbox{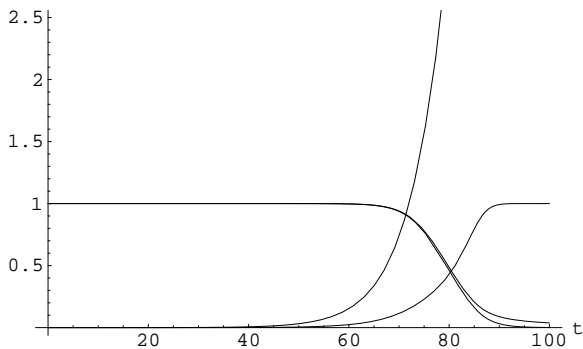}
\end{center}
\caption{Plots are for the time evolution of $T,~\dot T,~H$ and 
$\sqrt{8\pi G V / 3} $ presented in clockwise manner, with initial 
values  $H(0) = 1$, $T(0) =10^{-5}$. 
The value of $N_e$, which is the area under $H(t)$ curve in the 
plateau region, could easily 
be estimated to be close to 60. 
 $H$ vanishes at late time but $V(T)$ 
vanishes faster than $H$.} \label{fig1}
\end{figure}

\begin{figure}[!ht]
\leavevmode
\begin{center}
\epsfysize=5cm
\epsfbox{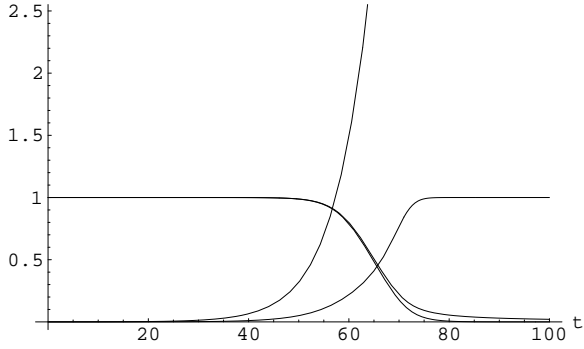}
\end{center}
\caption{For comparison these 
plots are for $T,~\dot T,~H$ and 
$\sqrt{8\pi G 
V / 3} $ but with initial values 
$H_0=1$ and  $T(0) =10^{-4}$. 
The value of $N_e$ could easily 
be estimated to be less than 60 in this case. $H$ vanishes at late time 
but faster than in the previous graph. } \label{fig2}
\end{figure}

\begin{figure}[!ht]
\leavevmode
\begin{center}
\epsfysize=5cm
\epsfbox{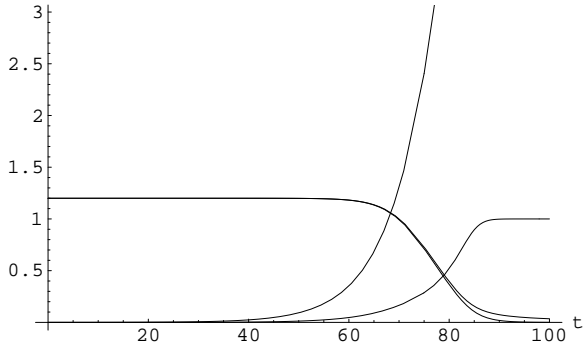}
\end{center}
\caption{Plots are for the time evolution of $T,~\dot T,~H$ and 
$\sqrt{8\pi G 
V / 3} $ in clockwise manner,  with higher initial value 
$H_0=1.2$, and $T(0)$ same as in second figure. 
The value of $N_e$ could easily 
be estimated to be  close to 70. }\label{fig3}
\end{figure}

The numerical results are plotted in figures \ref{fig1}, \ref{fig2} and 
\ref{fig3}. The time evolution of the quantities $T,~\dot T,~H$ and 
$\sqrt{8\pi G V / 3} $ could be found in these figures. The plots of 
$H$ and $\sqrt{8\pi G V / 3} $ overlap with each other in the plateau
region which characterises the inflationary period. 
Both $H$ and $V$ vanish at late time, but $V(T)$ 
vanishes faster than $H$. This is consistent with the bound we proposed in 
eq. \eqn{inte1}.
From fig.\ref{fig1}, it 
is found that taking the initial value $H(0)= 1 $ and the $T(0)\sim 
10^{-5}$ provides us with the number of e-folds during inflation close to 
$60$.
The duration of inflation is however very short and is around $60 (M_p)^{-1}$.
But it can be increased by taking smaller values of $H(0)$ and 
$T(0)$.
Also we observe that increasing the height of the tachyon potential  
increases
the number of e-folds as can be seen by comparing fig.\ref{fig2} and 
fig.\ref{fig3}.
From figures \ref{fig1} and \ref{fig2} we see that lower is the value of 
$T(0)$ longer is the duration of inflation which is on the expected lines.

\section{Conclusion}

We have studied rolling tachyon cosmological solutions in 
de Sitter space. The Friedman-Robertson-Walker solutions are taken to be
spatially flat and those are time-reversal non-symmetric. Only 
non-singular  configurations are allowed as the solutions
and rolling tachyon can give rise to
the required inflation, which can  be obtained by suitably choosing 
the initial conditions at the top of the tachyon potential. 
 For non-vanishing $\Lambda$, the matter dominated decelerating 
phase, with $a(t)\propto t^{2\over3}$, precedes the vacuum dominated 
accelerating expansion ($a(t)\propto e^{\lambda t}$) at the end of 
the evolution. However, we do not encounter the radiation dominated 
phase, where $a(t)\propto t^{1\over2}$.
We also find that cosmological constant of the universe has to be 
fine-tuned 
at the level of the effective action itself. In which case KKLT model 
\cite{kklt} of string
compactification can be a good starting point with 
$\Lambda_{KKLT}=\Lambda_{observed}$, indicating 
that string 
dynamics has to be responsible for such a small parameter as 
cosmological constant today. Interestingly, whatever the initial value of 
the Hubble constant at $t=0$, the  universe exits in vacuum dominated 
phase at the end of evolution.
We also find that there is an upper bound on 
the largeness of the compactification volume given as 
$$1<{v_0\over g_s^2}\le 10^{60}.$$

\leftline{\bf Acknowledgement:}
I would like to thank S. Roy for going through the draft and suggesting 
useful
changes and also bringing reference \cite{garousi} to my notice.

\end{document}